\documentclass[a4paper,twocolumn,aps,prd,nolongbibliography,superscriptaddress,showpacs,showkeys,amsmath,amssymb,floatfix,nofootinbib]{revtex4-1}
\usepackage{lmodern}
\usepackage{graphicx}

\usepackage[T1]{fontenc}
\usepackage[utf8]{inputenc}
\usepackage{color}
\usepackage[unicode=true,pdfusetitle,
 bookmarks=true,bookmarksnumbered=true,bookmarksopen=true,bookmarksopenlevel=1,
 breaklinks=false,pdfborder={0 0 0},backref=false,colorlinks=true]
 {hyperref}
\hypersetup{
 citecolor=blue,filecolor=blue,linkcolor=blue,urlcolor=blue}
\usepackage[normalem]{ulem}
\makeatletter

\pdfpageheight\paperheight
\pdfpagewidth\paperwidth

 \@ifundefined{textcolor}{}
 {%
   \definecolor{BLACK}{gray}{0}
   \definecolor{WHITE}{gray}{1}
   \definecolor{RED}{rgb}{1,0,0}
   \definecolor{green}{rgb}{0,0.7,0}
   \definecolor{BLUE}{rgb}{0,0,1}
   \definecolor{CYAN}{cmyk}{1,0,0,0}
   \definecolor{MAGENTA}{cmyk}{0,1,0,0}
   \definecolor{YELLOW}{cmyk}{0,0,1,0}
   \definecolor{SIENA}{rgb}{0.63,0.32,0.176}
 }

\newcommand{\doc}{\Delta\Omega_{\text{c}0}}

\renewcommand{\[}{\begin{equation}}
\renewcommand{\]}{\end{equation}}
\usepackage{array}
\makeatother

\begin{document}
\title{Using dark energy to suppress power at small scales} \thanks{Based in part on observations obtained with Planck (http://www.esa.int/Planck), an ESA science mission with instruments and contributions directly funded by ESA Member States, NASA, and Canada.}

\author{Martin Kunz}
\affiliation{D\'epartment de Physique Th\'eorique and Center for Astroparticle Physics,
Universit\'e de Gen\`eve, Quai E. Ansermet 24, CH-1211 Gen\`eve 4, Switzerland}
\affiliation{African Institute for Mathematical Sciences, 6-8 Melrose Road, Muizenberg,
Cape Town, South Africa}

\author{Savvas Nesseris}
\affiliation{D\'epartment de Physique Th\'eorique and Center for Astroparticle Physics,
Universit\'e de Gen\`eve, Quai E. Ansermet 24, CH-1211 Gen\`eve 4, Switzerland}

\author{Ignacy Sawicki}
\affiliation{D\'epartment de Physique Th\'eorique and Center for Astroparticle Physics,
Universit\'e de Gen\`eve, Quai E. Ansermet 24, CH-1211 Gen\`eve 4, Switzerland}

\begin{abstract}
The latest Planck results reconfirm the existence of a slight but chronic tension between the best-fit Cosmic Microwave Background (CMB) and low-redshift observables: power seems to be consistently lacking in the late universe across a range of observables (e.g.~weak lensing, cluster counts). We propose a two-parameter model for dark energy where the dark energy is sufficiently like dark matter at large scales to keep the CMB unchanged but where it does not cluster at small scales, preventing concordance collapse and erasing power. We thus exploit the generic scale-dependence of dark energy instead of the more usual time-dependence to address the tension in the data. The combination of CMB, distance and weak lensing data somewhat prefer our model to $\Lambda$CDM, at $\Delta\chi^2=2.4$. Moreover, this improved solution has $\sigma_8=0.79 \pm 0.02$, consistent with the value implied by cluster counts.
\end{abstract}
\maketitle

\section{Introduction}

The recent Planck cosmology results \cite{Ade:2013zuv,Ade:2015xua} provide stunning support for the $\Lambda$CDM ``standard model''
of cosmology. One of the few results that are not in quite as excellent agreement with the parameter constraints from the measurements of the anisotropies in the cosmic microwave background (CMB) is the determination of
the amplitude of cosmological perturbations at late times on small scales. This slight tension is most apparent when comparing to weak lensing measurements as provided by CFHTLenS, which prefers lower values of $\sigma_8$ (Ref.~\cite[Fig.~18]{Ade:2015xua}) and also the lower-than-expected cluster abundances (Ref.~\cite[Fig.~10]{Ade:2013lmv} and \cite{Ade:2015fva}). In addition, dark matter growth rates, proportional to $\sigma_8$, tend to fall on the low side of the concordance values, even if they are not inconsistent \cite[Fig.~16]{Ade:2015xua}. Although it is well possible that these discrepancies come from systematic effects in the analysis of the data in the low-redshift universe, it is also the case that late-universe measurements of $\sigma_8$ consistently imply a smaller value than that obtained by processing the initial amplitude of perturbations observed in the CMB through concordance gravitational collapse.

One possible resolution of this conflict is to suppress the clustering at low redshift through a new physical mechanism active only at late times. For example, heavy neutrinos could play such a role for clusters \cite{Wyman:2013lza, Battye:2014qga}. However, the latest data do not seem to support this \cite{Ade:2015xua}.

Somewhat more exotically, dark energy could be dynamical, changing the evolution of the universe at late times. However, the analysis in \cite[Fig.~4]{Ade:2015rim} shows  that distance data from baryon acoustic oscillations (BAO) constrains the usual quintessencelike models too much to significantly improve the possible fits to weak lensing data. More general models of modified gravity in the quasistatic limit that do not contain ghosts typically serve to increase growth rates \cite{Piazza:2013pua,Nesseris:2008mq,Nesseris:2009jf,Basilakos:2013nfa,Nesseris:2013jea,Nesseris:2011pc,
Nesseris:2013fca,Nesseris:2014mea}, although a temporary suppression around redshift $z \approx 1$ is also typical \cite{Perenon:2015sla} and could possibly be exploited with some tuning.

There is another possibility: dynamical dark energy could affect dark-matter clustering in a scale-dependent manner. The CMB is mostly a large-scale observable, while galaxy weak lensing has only been measured on small scales; clusters are also small-scale phenomena. A model where DE causes the dark matter to cluster slower at small scales than at large without significantly modifying the background expansion history could produce the desired phenomenology.

We propose a mechanism to achieve this: we exploit the \emph{dark degeneracy} \cite{Kunz:2007rk} to trade some of the dark matter for dark energy. Provided the two are similar enough, observables remain unchanged with respect to $\Lambda$CDM. We investigate a very minimal extension of the concordance cosmological model: we allow the dark energy to be a perfect fluid with constant pressure and a constant nonzero sound speed. This keeps the expansion history exactly the same as $\Lambda$CDM. Such a choice allows us to concentrate on investigating the effect of changing the properties of the perturbations of dark energy while keeping the background, which is already very well constrained, fixed.

If the dark energy has exactly zero sound speed, then the dark degeneracy also ensures that the linear perturbations in the metric are unchanged with respect to $\Lambda$CDM. Effectively, the dark energy component that replaces part of the dark matter also clusters like that missing dark matter. However, if the sound speed is nonzero then the dark energy clusters less than the missing dark matter on scales inside its sound horizon. We show that this model, as a result of this change in the behavior of DE at its sound horizon, indeed goes some way to relieve the tension between the early and late universe and is favored over $\Lambda$CDM when the combination of CMB and weak lensing data are used. Moreover, the best-fit model has a lower value of $\sigma_8$ which should help with the tension with cluster number counts.

\section{Dark-Energy Model}

All measurements of the cosmological background expansion history, such as baryon acoustic oscillations (BAO) or supernovae (SNe), or even cosmic chronometers, only measure relative distances or times, which are integrated quantities depending only on the dimensionless Hubble parameter $H(z)/H_0$, where $H_0$ is the Hubble constant. Thus, provided that the equation of state of the total dark sector be kept constant (i.e.~the sum of CDM and DE), distance measurements are completely degenerate to any changes in relative composition between the two species. This fact was named the \emph{dark degeneracy} in Ref.~\cite{Kunz:2007rk} and it implies that background measurements are incapable of measuring the cold dark matter (CDM) fraction $\Omega_{\text{c}0}$ absent a choice of parametrization for $w_X$. Measurements of large-scale structure, such as the CMB anisotropies, can break this degeneracy, but require a model for the behavior of dark-energy perturbations to be specified \cite{Amendola:2012ky, Motta:2013cwa}. One should really think of the density fraction of DM as a perturbation variable and not a background parameter \cite{Bellini:2014fua}. While $w_X$ is frequently parametrized as constant and the best-fit cosmology prefers values close to $w_X=-1$ with a good determination of $\Omega_{\text{c}0}$ \cite{Ade:2015xua}, the indubitable fact is that our universe seems to be close to one which has a constant \emph{pressure} at late times, making a DE component with a constant energy density just one of the possibilities. As we will show here, this is not just an academic discussion.

Our only modification to the concordance model is to use a perfect-fluid dark energy with a constant (rest-frame) sound speed $c_\text{s}$ and a particular designer choice of equation of state $w_X$ that ensures that the background expansion history \emph{exactly} mimics $\Lambda$CDM (i.e.~our DE has constant \emph{pressure}).

\subsection{Background}
In order to maintain a fixed expansion history, we pick the density fraction of dark energy $\Omega_X$ such that its sum with the CDM density fraction remains the same as the sum of the density fraction of the cosmological constant and some reference CDM density $\Omega_{\text{c}0}^\text{Planck}$ in $\Lambda$CDM. We will refit the data for the values of all the parameters in the paradigm of our extended model, but for the moment, one should think of it as the standard Planck value for $\Omega_{\text{c}0}$,
\[
\Omega_X(a) + \Omega_\text{c}(a) = \Omega_\Lambda(a) + \Omega_\text{c}^{\rm Planck}(a) \,,
\]
where $a$ is the scale factor. This relation implies that the equation of state of the DE satisfies
\begin{equation}
1+w_X(a) = \frac{\Delta\Omega_{\text{c}0}}{\Delta\Omega_{\text{c}0} + \Omega_{\Lambda0} a^3} \label{eq:wX}\,,
\end{equation}
where $\Delta\Omega_\text{c} \equiv \Omega_c^{\rm Planck} - \Omega_c$ and where the subscript $0$ denotes the value today. $\Delta\Omega_\text{c0}$ is the only new parameter introduced at the level of the background and it replaces the more typical constant parametrization for $w_X$.

Generically, a violation of the null energy condition leads to instabilities. These are inescapable for NEC-violating perfect fluids (either ghost or gradient instabilities) \cite{Dubovsky:2005xd}, but may even exist in more general cases \cite{Sawicki:2012pz}. We thus only consider $\Delta\Omega_\text{c}>0$, i.e.~$w_X>-1$.

We note here that such choices for $w_X$ as Eq.~\eqref{eq:wX} are badly captured by the prevailing parametrizations in use in the community, such as a constant or the CPL parametrization \cite{Chevallier:2000qy,Linder:2002et}, which is one of the origins of our results. It is much more similar to models with a transition in the equation of state (e.g.~Ref.~\cite{Bassett:2002qu}) or Early Dark Energy (EDE) parametrizations (e.g.~Ref.~\cite{Pettorino:2013ia}), since the DE component tracks the CDM before the acceleration era. Other models with similar $w(z)$ include quintessence coupled to neutrinos \cite{Amendola:2007yx} or the generalized Chaplygin gas (GCG) \cite{Kamenshchik:2001cp}.

The evolution of the background is essentially missing all information about the microphysical content of the DE model, which is defined through the behavior of perturbations. For example, modeling the GCG perturbations in various ways gives rise to completely different constraints \cite{Sandvik:2002jz,Wang:2013qy}. We focus on defining the evolution of perturbations next.

\subsection{Linear properties}\label{sec:linprops}

We start off with a Friedmann-Robertson-Walker (FRW) metric with small scalar perturbations, with the gauge fixed to be Newtonian:
\begin{equation}
\mathrm{d}s^2 = -(1+2\Psi)\mathrm{d}t^2 + a^2(t)(1-2\Phi)\delta_{ij}\mathrm{d}x^i\mathrm{d}x^j\,,
\end{equation}
keeping only the scalar perturbations.

By construction, the scalar sector of the linear perturbation equations for the dark energy takes the standard form for a perfect fluid which we do not give here (see e.g. \cite{Hu:1998kj,dePutter:2010vy,Sapone:2009mb}). Since the background expansion history mimics $\Lambda$CDM, the pressure of the dark energy is constant and therefore the adiabatic sound speed $c_\text{a}^2=0$ exactly. Therefore the only new parameter that we need to specify is the sound speed $c_\text{s}$ which we take to be constant and to lie in the range $0<c_\text{s}^2 \leq 1$. This is the second and final new parameter of our model. We can define the (physical) Jeans scale as
\begin{equation}
	k_\text{J}(z) \equiv \frac{H(z)}{(1+z)c_\text{s}}\,,
\end{equation}
which we will demonstrate separates two different regimes of evolution for perturbations.

Provided that the dark energy consists of a single degree of freedom, and therefore has no additional internal freedom, the dynamics of the perturbed cosmology can be rewritten in a manner much more illustrative than the usual presentation with fluid variables (see \cite[Eq.~(21)]{Motta:2013cwa}): one combines the various constraints given by the Einstein equations to eliminate the DE fluid variables and to form a single equation of motion for the Newtonian potential which is coupled to the other matter species,
\begin{align}
\ddot\Phi + & 3H\dot\Phi + H\dot\Psi   +    \left(  3H^2  +  2\dot{H}   \right) \Psi + c_\text{s}^2   \frac{k^2}{a^2} \Phi= \label{eq:phievol}\\
& = -c_\text{s}^2  \left( \frac{\rho_\text{m}\delta_\text{m}}{2}  -  \frac{3}{2} (\rho_\text{m}+p_\text{m})  H v_\text{m}  \right) +  \frac{\delta p_\text{m}}{2}   \notag\\
\Psi-\Phi  &   =   p_\text{m}\pi_\text{m}
\end{align}
where the fluid variables with subscript ``m" describe the total perturbation of the matter sector, including radiation, baryons, CDM and neutrinos, but not the DE. This system is closed by supplying the evolution equations for the total matter energy-momentum tensor, whether by assuming the matter be a collection of fluids, or by solving some sort of Boltzmann hierarchy.

Equation~\eqref{eq:phievol} should be contrasted with its version in $\Lambda$CDM:
\begin{equation}
\ddot\Phi + 3H\dot\Phi + H\dot\Psi   +    \left(  3H^2  +  2\dot{H}   \right) \Psi = \frac{\delta p_\text{m}}{2}   \label{eq:phievol-lcdm} \,.
\end{equation}
There are three important differences between Eqs.~\eqref{eq:phievol} and \eqref{eq:phievol-lcdm}:
\begin{itemize}
	\item the presence of a sound-horizon (Jeans) term $c_\text{s}^2\frac{k^2}{a^2}\Phi$;
	\item the sourcing of the potential by the matter perturbations with a coupling $c_\text{s}^2$;
	\item the potential $\Phi$ in our model is in fact an independent degree of freedom with its own initial conditions.%
	\footnote{Note that in tracker models of DE, the isocurvature modes decay away \cite{Malquarti:2002iu}.}
	In $\Lambda$CDM, it is completely constrained and not independent.
\end{itemize}
Note that the cosmological horizon does not appear as a relevant scale here at all; its sound horizon is the only scale relevant for the DE.
\\

At scales larger than the DE sound horizon, $k\ll k_\text{J}$, the quasistatic approximation for the DE cannot be used \cite{Sawicki:2015zya}. The Jeans term on the left-hand side of Eq.~\eqref{eq:phievol} is negligible and the difference with respect to the $\Lambda$CDM case is the new source term coupled with $c_\text{s}^2$ on the right-hand side. It affects the evolution of the potentials whenever it is comparable to the pressure perturbation term $\delta p_\text{m}$ coming from radiation, essentially providing a new early integrated Sachs-Wolfe contribution. For example, around the time of decoupling, the CDM and radiation density fractions are comparable, so the new term is negligible when $c_\text{s}^2\ll 1/3$. For such DE sound speeds, the evolution of perturbations in the metric is the same as in $\Lambda$CDM, irrespective of the value of $\Delta\Omega_\text{c}$, i.e.\ how much CDM has been removed and replaced with the DE.

This means that, outside of its sound horizon, the DE clusters just like CDM since it cannot react causally to create a pressure to arrest the dustlike collapse along geodesics. The fact that quintessence tracking models behave in this way at large scales was shown in Ref.~\cite{Malquarti:2002iu} and this is also one of the implications of the separate-universe approach \cite{Bertschinger:2006aw}. Essentially, the part of the DE energy density that is tracking contributes in the same way as the CDM dust at these scales and therefore the model looks like $\Lambda$CDM there.
\\

At scales inside the Jeans scale, $k \gg k_\text{J}$, the Jeans term makes the evolution equation \eqref{eq:phievol} stiff, leading to the solution usually known as the quasistatic solution,
\begin{equation}
\frac{k^2}{a^2}\Phi\approx -\frac{\rho_\text{m}\delta_\text{m}}{2}  +  \frac{3}{2} (\rho_\text{m}+p_\text{m})  H v_\text{m} +\frac{\delta p_\text{m}}{2c_\text{s}^2} \,. \label{eq:Poisson}
\end{equation}
When the pressure is irrelevant, this recovers the standard Poisson equation of $\Lambda$CDM, with the gravitational potential driven purely by the comoving density fluctuations of matter. At these scales, the DE reacts causally and sets up a pressure profile to prevent its clustering. The gravitational potential is sourced only by the standard species. Thus picking a $\Delta\Omega_\text{c}>0$ to replace a part of CDM will reduce the depth of the potentials and therefore also the growth rate on scales inside the DE Jeans length.

When the total pressure perturbation is not negligible compared to the total density perturbation, e.g.~at decoupling, the final term in Eq.~\eqref{eq:Poisson} could in principle contribute for small DE sound speeds. On the other hand, we have to bear in mind that the angular scale of the DE Jeans scale in the CMB is proportional to the DE sound speed. We can use the Limber approximation to estimate that the multipoles at  which the approximation  \eqref{eq:Poisson} is valid is
\begin{equation}
\ell\gg \ell_\text{Jeans} \sim \ell_\text{peak} \frac{\sqrt{3}}{c_\text{s}}\,, \label{eq:l-Jeans}
\end{equation}
where $\ell_\text{peak}\approx200$ is the multipole at which the first CMB peak is centered and therefore the solution \eqref{eq:Poisson} is only possibly relevant to the observed CMB anisotropies for $c_\text{s}^2>10^{-2}$. 

\begin{figure}
	\includegraphics[width=1\columnwidth]{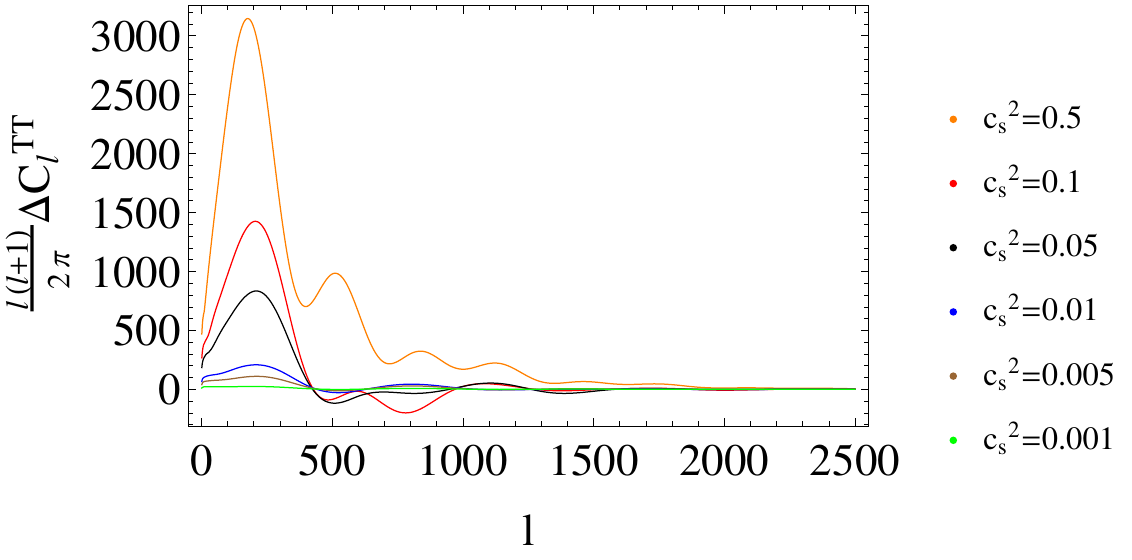}
	\caption{\label{fig:cmb-cs2-effect}The effect on the unlensed CMB anisotropy TT power spectrum of replacing roughly a third of the CDM ($\Delta\Omega_\text{c0}=0.1$) with our DE, as a function of its sound speed $c_s^2$. For clarity we show the difference between the power spectra for a given $c_s^2$ and the $\Lambda$CDM model ($c_s^2=0$), ie $\Delta C_l^{TT}=C_l^{TT}(c_s^2)-C_{l}^{TT}(c_s^2=0)$. A large sound speed increases the amplitude of the peaks, since the DE and radiation Jeans horizons are close and the decay of perturbations is much more rapid than in $\Lambda$CDM. By comparing to a perfect Planck forecast with no lensing, we have verified that the unlensed power-spectrum stops being sensitive to DE for sound speeds as large as $c_\text{s}^2\sim 10^{-3}$.}
\end{figure}

We illustrate the effect of our model on the CMB TT power-spectrum in Fig.~\ref{fig:cmb-cs2-effect}, where it can be seen that already for $c_\text{s}^2=0.001$, the modification of the $\Lambda$CDM curve from the terms discussed above is negligible when the effects from the late universe are not included. This can be verified by comparing a perfect Planck forecast with no late-time effects included for $\Lambda$CDM with our model.
\\

\begin{figure}[tb]
	\includegraphics[width=1\columnwidth]{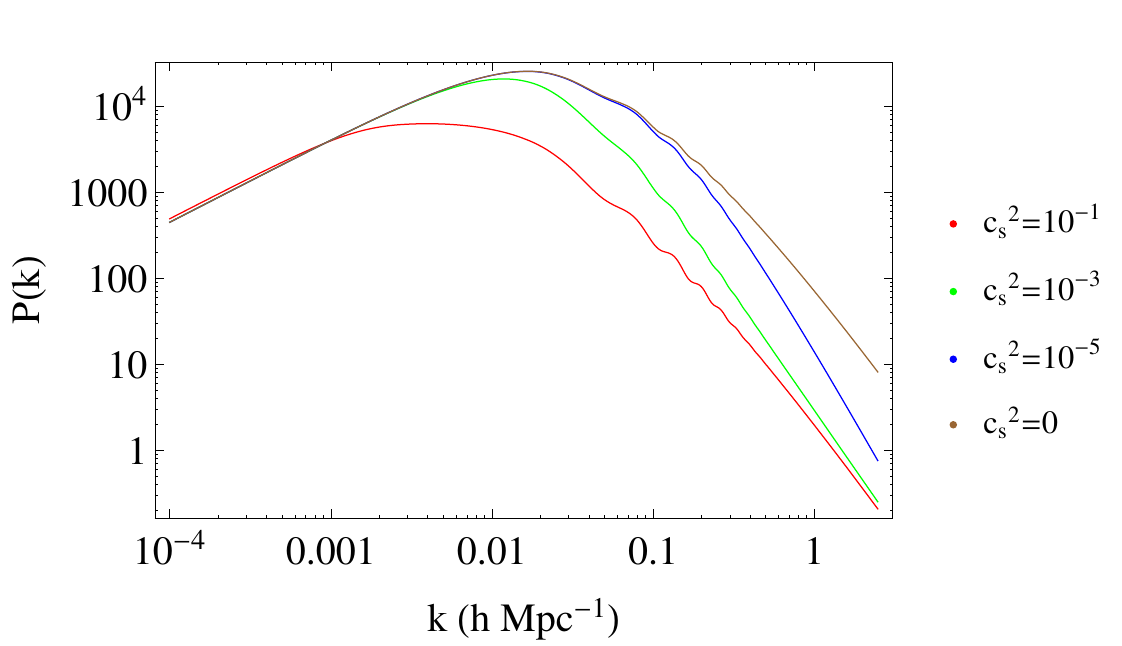}
	\caption{\label{fig:Pk-cs2-effect} The effect on the CDM power spectrum at $z=0$ of replacing roughly a third of the CDM ($\Delta\Omega_\text{c0}=0.1$) with our dark energy. For $k\gg H_0/c_\text{s}$, the DE does not cluster and causes the CDM growth rate to decrease after the mode crosses the DE sound horizon. This suppresses the power spectrum at small scales in a scale-dependent manner. Scales larger than the sound horizon are not affected and match $\Lambda$CDM.}
	\includegraphics[width=1\columnwidth]{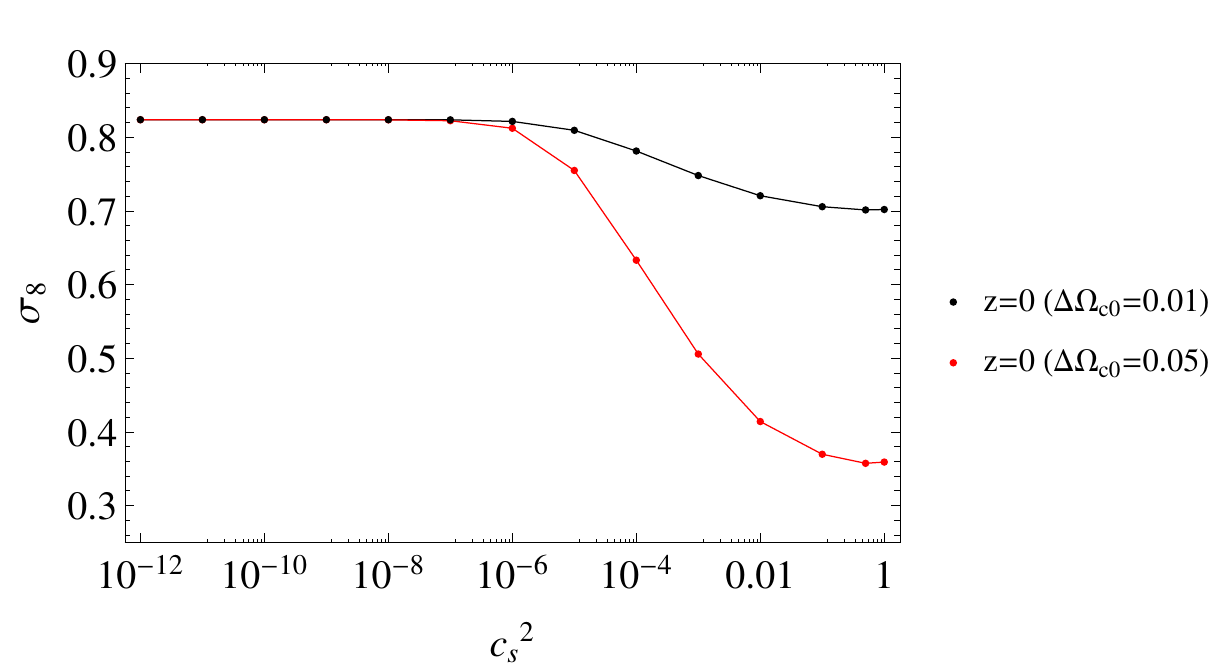}
	\caption{The effect on the normalization of the CDM power spectrum, $\sigma_8$, at $z=0$ of replacing part of the CDM with our dark energy. For $c_\text{s}^2\lesssim 10^{-6}$, the power spectrum is significantly suppressed only on scales smaller than 8~Mpc and therefore $\sigma_8$ is not affected by such low sound speeds. As $c_\text{s}^2$ increases, the power spectrum is erased on increasingly large scales, reducing $\sigma_8$.\label{fig:s8-cs2-effect}}
\end{figure}

Turning to the late universe, as the mode crosses the DE sound horizon, the standard $\Lambda$CDM-like solution for the potential $\Phi$ valid super-Jeans is modified to \eqref{eq:Poisson} as the DE perturbations are erased. This then modifies the growth function for that mode at all subsequent times, with the matter perturbation evolving according to
\[
	\ddot\delta_\text{m} + 2H\dot{\delta}_\text{m}-\frac{3}{2}H^2(1-\Omega_X)\delta_\text{m}=0\,.
\]
This is essentially the same equation as in the $\Lambda$CDM case, apart from the fact that even during matter domination $\Omega_X\approx\Delta\Omega_\text{c0}/\Omega_\text{c0}\neq0$, which changes the growing mode evolution during matter domination from $\delta_\text{m}^{\Lambda\text{CDM}}\propto a$ to
\[
	\delta_\text{m}\propto a^p\,,\quad p\approx 1-\frac{3\Omega_X}{5}\,,
\]
for $\Omega_X\ll1$. This means that the matter growth functions at some redshift $z$ during matter domination receives a scale-dependent correction:
\[
	G(k,z)\propto\left( \frac{k_\text{J}(z)}{k} \right)^{\frac{3}{5}\Omega_X} (1+z)^{-1}\,, \qquad k>k_\text{J}\, \label{eq:growth}
\]
where $z$ is the redshift of observation. Therefore, even for a fixed amount of dark energy $\doc$, a larger sound speed means that the pivot of the suppression effect lies at smaller $k$ and therefore the reduction in power is higher for any particular mode inside the sound horizon (see Fig.~\ref{fig:Pk-cs2-effect} for the effect on the CDM power spectrum or Fig.~\ref{fig:s8-cs2-effect} for an equivalent presentation of how the CDM power spectrum amplitude $\sigma_8$ today depends on the amount and sound speed of our DE).

We remind the reader that the CDM and gravitational potential power spectra are not affected for $k<k_\text{J}$, since the dark energy is clustering and the potential has the full depth as in $\Lambda$CDM.%
\footnote{At very late times, the comoving horizon shrinks as a result of the acceleration. Thus modes can in principle exit the DE sound horizon and have their growth increase back to $\Lambda$CDM rates.}%

In principle, the amplitude of $\sigma_8$ is closely related to the rate of formation and mass function of halos. However, we should note here that two different $\sigma_8$'s can be defined: one given by the autocorrelation of the CDM density contrast, while the other defined from the autocorrelation of the gravitational potential. In the $\Lambda$CDM, $c_\text{s}^2\rightarrow 0$ limit, they are identical. But whenever the DE Jeans length lies at scales larger than 8~comoving Mpc, the two are different, with the former larger. On one hand, the forming clusters only know about the gravitational potential, since the CDM does not interact in any way but through gravity. On the other hand, the clusters form in configuration space and locally only the mass in the CDM perturbations is available to form them. Which of the definitions is more appropriate requires further study outside of the scope of this work and we will use the more usual CDM definition henceforth.

\begin{figure}
	\includegraphics[width=1\columnwidth]{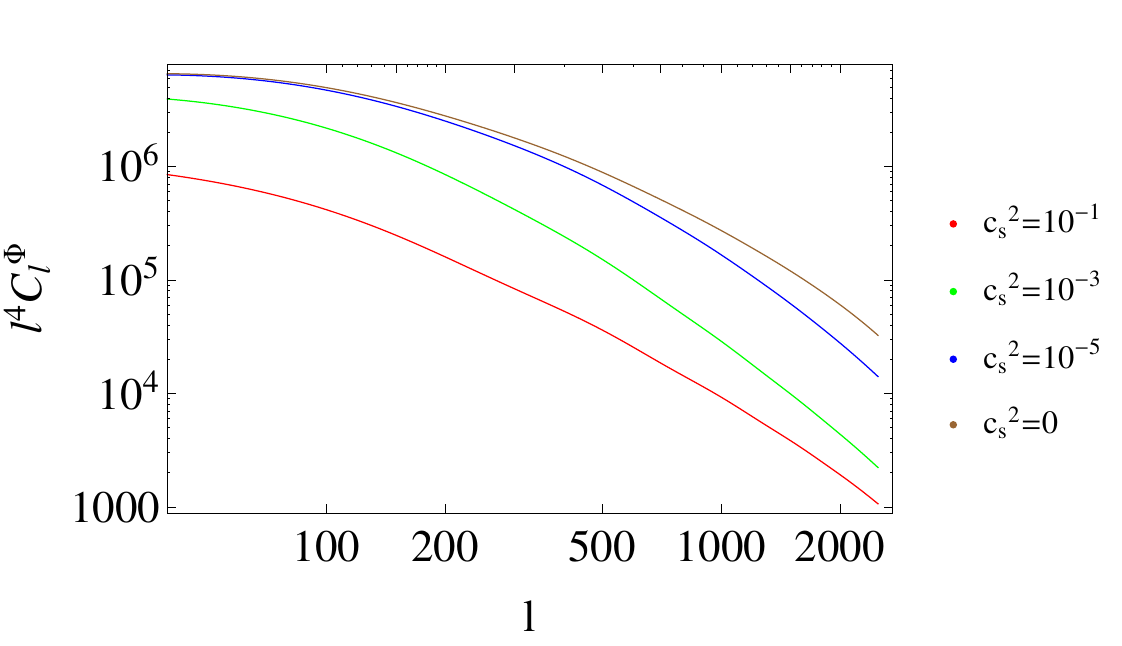}
	\caption{\label{fig:lenspot}The effect of replacing a third of the CDM ($\Delta\Omega_\text{c0}=0.1$) on the projected gravitational lensing potential power spectrum $l^4 C_l^\Phi$ as a function of the DE sound speed squared $c_\text{s}^2$. Large values of the sound speed erase the contribution to the potential from the DE even at the largest scales. As the sound speed is decreased, the pivot of the damping moves to higher multipoles, eventually giving the same lensing effect as $\Lambda$CDM. The Planck temperature power spectrum constrains the amplitude of the lensing signal to about $10\%$, which allows CMB lensing to break the dark degeneracy for $c_\text{s}^2\gtrsim 10^{-5}$.}
\end{figure}

The scale-dependent reduction of the amplitude of the gravitational potential affects lensing. We plot the lensing potential in Fig.~\ref{fig:lenspot}, showing that the suppression is very strong for large sound speeds and for large $\doc$. The effect of lensing on the CMB was detected by Planck by two methods: the smoothing of the power spectrum at small scales and using the trispectrum; and it matches that predicted by $\Lambda$CDM closely (although the power-spectrum method sees a 2$\sigma$ excess of lensing in the context of $\Lambda$CDM) \cite{Ade:2015xua}. As we have already mentioned and show in Fig.~\ref{fig:cmbtri}, the CMB at decoupling is only mildly sensitive to DE sound speeds. It is the reduction of the CMB lensing effect in the late universe that breaks the degeneracy between the DE and CDM when $c_\text{s}^2>10^{-5}$.

Reference~\cite{Bean:2003fb} proposed that the integrated Sachs-Wolfe effect could be used to constrain the sound speed of dark energy and shows that WMAP data imply $c_\text{s}^2<0.04$. Since the evolution of gravitational potentials at scales $k\ll k_\text{J}$ in our model is purely determined by the evolution of the cosmological background, the integrated Sachs-Wolfe (ISW) effect on the CMB anisotropies also matches that of $\Lambda$CDM for a low-enough sound speed and this does not provide an interesting constraint. Similarly, cross-correlations of ISW with galaxy clustering are unlikely to constrain this model beyond some minimal sound speed \cite{Hu:2004yd}. For lower sound speeds, the above probes only test the expansion history.%
\footnote{As was argued in Ref.~\cite{Bertschinger:2006aw}, the presence of gravitational slip changes the evolution of the potential at large scales. Thus the late-time ISW effect could still be used to constrain it, even for a fixed expansion history, and therefore test for modifications of gravity \cite{Saltas:2014dha}.} %
The novelty of Planck with respect to WMAP in this context is that CMB lensing provides new constraints which push the limits on $c_\text{s}^2$ much further than large-scale measurements.
\\

The picture we have therefore built up is that, for small enough DE sound speeds $c_\text{s}$, the dark energy perturbations behave as dust outside of their Jeans scale, making up for the removed CDM and giving essentially the same predictions as $\Lambda$CDM. Inside the Jeans scale, the DE reacts and develops a pressure which arrests its collapse, allowing only the CDM to cluster. This reduces the growth rate at small scales when compared to large, introducing a new scale dependence in all the observables. In the limit $c_\text{s}\rightarrow 0$, the dark energy behaves as dust at all scales, and therefore is completely degenerate with $\Lambda$CDM, despite its equation of state (see e.g.~the Angel Dust model of \cite{Lim:2010yk} or mimetic dark matter of \cite{Chamseddine:2013kea, Chamseddine:2014vna}).

As a result, the constraints that we obtain from the CMB are always an upper bound on the DE sound speed. Late-universe measurements will match the $\Lambda$CDM results if they are performed at scales larger than $k_\text{J}$ and will differ if they are centered on smaller scales. For very low DE sound speeds, there is total degeneracy. The scale-dependent growth rate resulting from the Jeans horizon allows us to alleviate some of the tension of CMB measurements when the sound speed is dialed up.

It is important to stress that at no point is the effective Newton's constant, describing the response of the gravitational potential to the CDM, smaller than its standard value. Deep inside the Jeans scale, they are exactly equal. Thus we have achieved a reduction of the growth rate without introducing some sort of gravity-like repulsive interaction which would in all likelihood be pathological.

\subsection{Microscopic model}

We have purposefully de-emphasized the discussion of the precise model of dark energy which might produce such phenomenology as above. Essentially any DE/MG model which comprises a single degree of freedom and has an energy-momentum tensor of perfect-fluid form gives this phenomenology when the solution has the equation of state \eqref{eq:wX} and a constant sound speed for the perturbations around it. Specifying the model allows for a calculation of any nonlinear effects and for an estimate of the domain of validity of our linear model.

As an example, the class of k-\emph{essence} models \cite{ArmendarizPicon:2000dh} provides for such behavior. The action is given by
\[
S_\phi = \int \mathrm{d}^4\! x \sqrt{-g} K(X,\phi) \, , \quad X \equiv - \frac{1}{2} g^{\mu\nu} \partial_\mu\phi \partial_\nu\phi \, .
\]
The equation of state and sound speed by this dark energy are given by
\[
	w=\frac{K}{2XK_{,X}-K}\,,\qquad c_\text{s}^2 = \frac{K_{,X}}{K_{,X}+2 X K_{,XX}}
\]
and given some initial conditions for $\phi,\dot{\phi}$ a suitable $K$ can be found by integrating the above. An appropriate choice could in fact be
\[
K=V(\phi)+M^4(X/M^4)^\frac{(1+c_\text{s}^2)}{2c_\text{s}^2}
\label{eq:action}
\]
whenever the sound speed is small $c_\text{s}^2\ll 1$ and where $V(\phi)$ is some sufficiently slowly varying potential, $M$ is a constant mass scale.

Another way of thinking about this model, which is completely equivalent at the level of this paper, is that we are introducing a subcomponent of the dark matter which exists in a condensate with a nonvanishing sound speed \cite{Sawicki:2013wja}. In such an interpretation, the acceleration is still being driven by a cosmological constant.

One should, however, be mindful when taking such scalar-field superfluid models literally: the behavior of perturbations when nonlinear deviates from standard fluids, since the scalar field cannot carry vorticity and therefore cannot virialize. The perturbations in the DE can become nonlinear when $c_{\text{s}}^2\lesssim 10^{-6}$ \cite{Batista:2013oca}. At the same time, for sound speeds $c_\text{s}^2\lesssim 10^{-5}$ the pressure support that the DE can provide is never quite enough to arrest the collapse and such a fluid would continue to collapse either until the breakdown of the effective description or would form a black hole.

\section{Results}\label{sec:results}

For our analysis we have slightly modified the CAMB/CosmoMC public codes \cite{Lewis:1999bs,Lewis:2002ah} to properly model the evolution of a perfect fluid with a time-dependent equation of state \eqref{eq:wX} and a constant sound speed $c_\text{s}$. Unless stated otherwise, we only let the parameters $\Omega_{c0}h^2, \doc, c_\text{s}^2$ and $\theta_\text{MC}$ vary, keeping the others fixed to their $\Lambda$CDM Planck 2013 best-fit values \cite{Ade:2013zuv}. Also, we should mention that we scan the parameter space in terms of $\log_{10}(c_s^2)$ instead of simply $c_s^2$, in order make it easier for the MCMC code to sample the subtle effects of the sound speed and recover the degeneracy for low $c_s^2$. For the final combined analysis, we have freed all parameters.

Whenever possible, we have tried following the relevant parts of the analysis of \cite{Ade:2015xua}, although we were only able to use the Planck likelihoods from the 2013 data release \cite{Ade:2013zuv}. We first discuss the constraints from the different data sets in turn and then perform a combined analysis in Sec.~\ref{sec:combined}.

In order to provide a reasonable representation of the degeneracies, we are always including distance data together with each of the perturbation-related data sets. Therefore, we have included the BAO measurements from CMASS and LOWZ of Ref.~\cite{Anderson:2013zyy}, the 6DF measurement from Ref.~\cite{Beutler:2011hx}, the MGS measurement from Ref.~\cite{Ross:2014qpa} and the Union 2.1 SNe Ia catalog from Ref.~\cite{Suzuki:2011hu}, all readily available in the CosmoMC code. We do not include any measurements of the Hubble constant $H_0$, apart from a uniform prior $0.4\leq h \leq 1.0$.

The setup most similar to ours was previously investigated in \cite{dePutter:2010vy}, albeit the parametrizations for the equations of state of DE used were different. At the time, WMAP did not provide strong constraints on the DE sound speed. As we show here, this is no longer the case in the context of a fixed background expansion history.

\subsection{Constraints from full CMB\label{sec:cmbl}}

The Planck papers on cosmology \cite{Ade:2015xua} and on dark energy and modified gravity \cite{Ade:2015rim} demonstrated that $\Lambda$CDM is a good fit to the data. Given the discussion in Sec.~\ref{sec:linprops}, we expect to recover a perfect degeneracy between $\Omega_\text{c0}$ and $\doc$ when the sound speed is sufficiently low.

For this fit we are using both the CMB temperature and low-multipole polarization power spectra from Planck including of course the lensing of the CMB. We have not included the lensing information extracted from the temperature trispectrum.

\begin{figure*}[tb]
\includegraphics[width=0.95\textwidth]{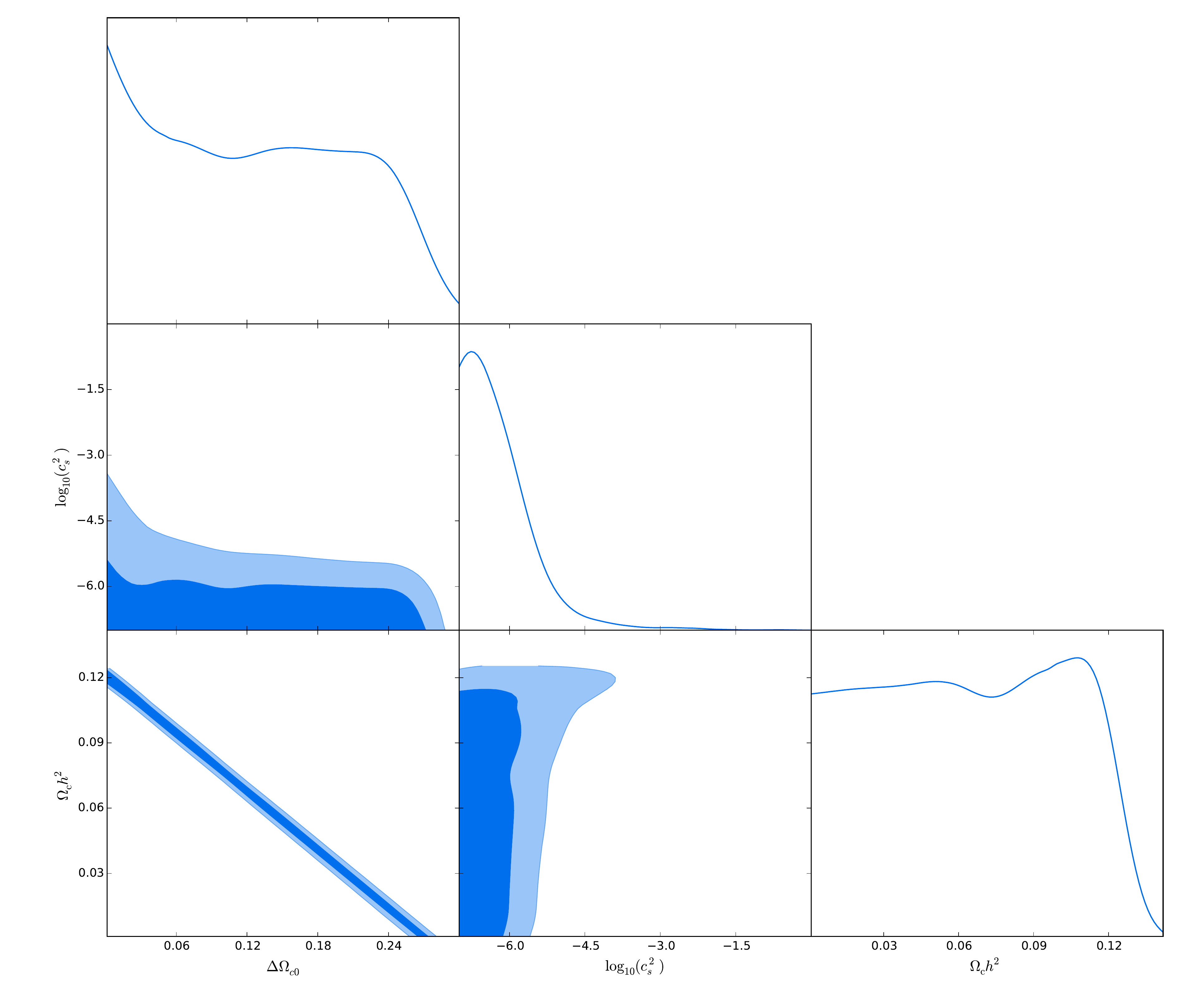}
\caption{The constraints on $\Omega_\text{c0}$, $\Delta\Omega_\text{c0}$ and $c_s^2$ from the CMB anisotropy power spectrum (including CMB lensing) and distance data (SNe+BAO). Provided the sound speed be $c_\text{s}^2<10^{-5}$, the model is degenerate with $\Lambda$CDM for these data. Larger DE sound speeds are allowed if the admixture of DE is sufficiently small. The constraint on the sound speed is essentially completely determined by the effect of CMB lensing in the late universe on the observed CMB power spectrum, and not on the shape of the CMB power spectrum at decoupling.\label{fig:cmbtri} }
\end{figure*}

As expected, Fig.~\ref{fig:cmbtri} shows that CMB anisotropies constrain only the sum $\Omega_\text{c0}+\doc=0.271 \pm 0.004$ whenever $c_\text{s}^2\lesssim 10^{-5}$. For small admixtures of DE, much higher sound speeds are also allowed. The CMB does not show any preference for having this extra component, but it cannot rule out its existence either.

As we have previewed already in Sec.~\ref{sec:linprops}, this constraint on $c_\text{s}^2$ actually comes mainly from CMB lensing, which takes place in the late universe. We have generated a perfect Planck forecast without lensing for $\Lambda$CDM and our model and have found that the dark degeneracy cannot be broken by the CMB power spectrum for $c_\text{s}^2\gtrsim 10^{-3}$. The CMB at decoupling is not very sensitive to clustering properties of the dark matter.

\subsection{Weak lensing shear\label{sec:WL}}

As was noted by Planck \cite{Ade:2015xua, Ade:2015rim}, the data from weak lensing shear are slightly incompatible with the Planck $\Lambda$CDM analysis: the preferred amplitude of the lensing potential found by CFHTLeNS at angular scales $1.5'<\theta<37.9'$ is lower than that expected when the initial normalization of perturbations implied by the CMB $A_\text{S}$ is extended to smaller scales and evolved forward in $\Lambda$CDM \cite{Heymans:2013fya}. If $A_\text{S}$ is kept constant at the Planck best-fit value, and the distance measurements are in the likelihood, the optimal solution for WL has a lower density fraction $\Omega_\text{c0}+\doc=0.244 \pm 0.084$ and a reduced amplitude for the gravitational potential sourced by a matter distribution with $\sigma_8 = 0.761 \pm 0.024$. Provided that the sound speed of our DE $c_\text{s}^2\lesssim 10^{-7}$, which ensures that the data lie at scales outside of the DE Jeans horizon, the WL data cannot break the dark degeneracy.

However, a new, equally good, solution for the lensing data appears in the presence of a small amount of our DE ($\doc=0.022 \pm 0.016$) with a large sound speed, $c_\text{s}^2\gtrsim 2.5\cdot 10^{-3}$ which also allows for a  much larger value of the CDM density fraction $\Omega_\text{c0}=0.265 \pm 0.055$. In such a scenario, the DE clusters only at largest scales and therefore the growth rate of the CDM is reduced according to Eq.~\eqref{eq:growth}. This reduces the amplitude of the lensing potential at late times despite the large $\Omega_\text{c0}$. We have verified that the existence of this solution does not strongly depend on the nonlinear completion used for the matter power spectrum: just as Planck, for the fits in this paper we use Halofit with the parameters of Takahashi et al.~\cite{Takahashi:2012em}. However, we have verified that the constraints are not biased by comparing them with the ones obtained using the original Halofit parameters \cite{Smith:2002dz} and even by switching off the nonlinear correction to the matter power spectrum completely. We have also used the correlation functions from the alternative CFHTLeNS analysis of Ref.~\cite{Kilbinger:2012qz}, verifying that it would not significantly influence our results.

\begin{figure*}
\includegraphics[width=0.95\textwidth]{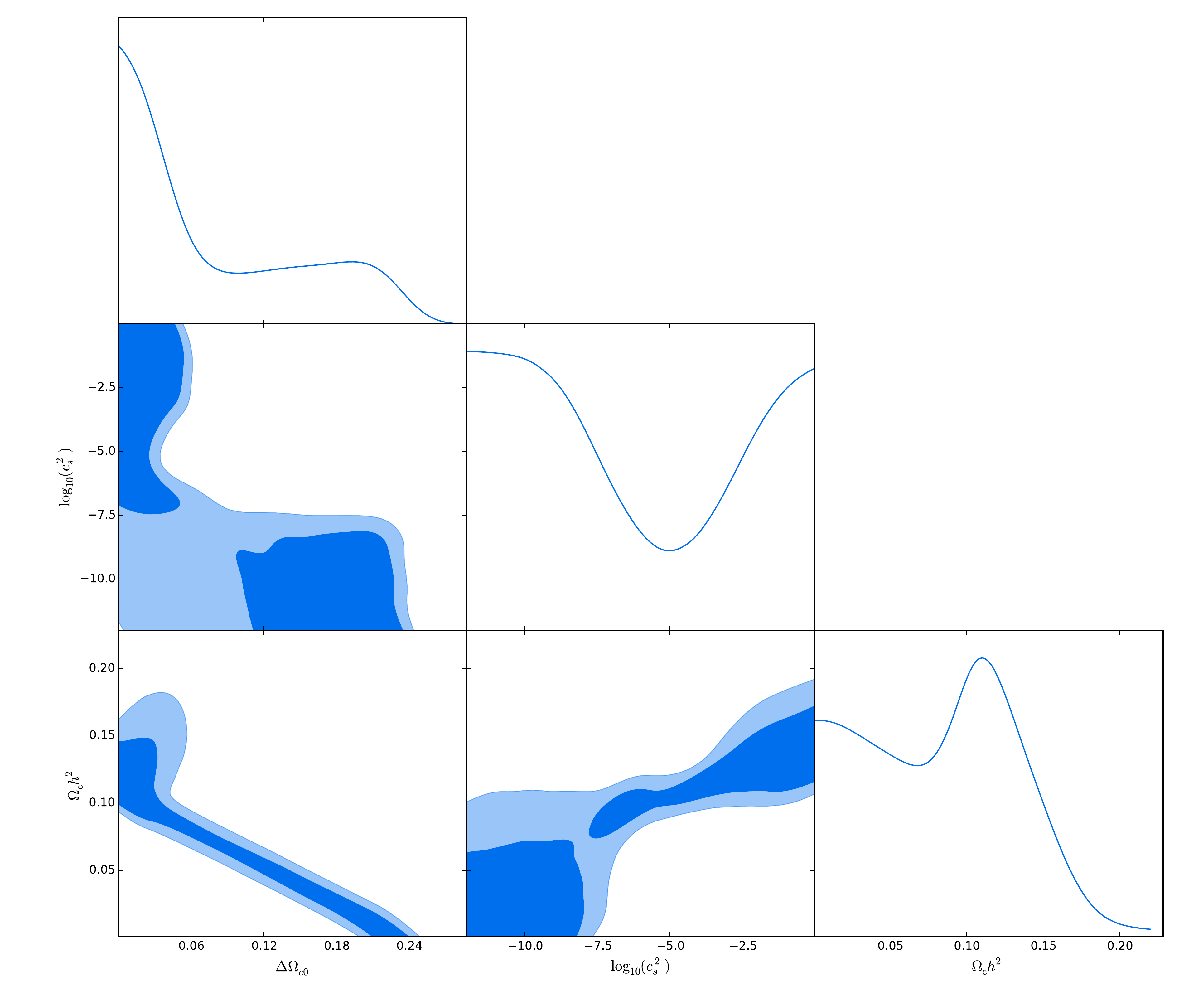}
\caption{\label{fig:wl} The constraints on $\Omega_\text{c0}$, $\Delta\Omega_\text{c0}$ and $c_s^2$ from CFHTLeNS weak-lensing data and distance data (SNe+BAO). WL data prefer a lower amplitude of perturbations than the CMB. Since we have fixed the initial inflationary amplitude to the Planck $\Lambda$ best-fit value, the solution achieves a good fit by reducing the preferred value of CDM density fraction $\Omega_\text{c0}$ compared to the CMB in figure~\ref{fig:cmbtri}. The dark degeneracy is only recovered for sound speeds $c_\text{s}^2\lesssim 10^{-7}$, compared to $10^{-5}$ in the case of the CMB, since the weak lensing data lie at much smaller scales than the CMB lensing. \\
The WL data also permit an alternative, equally well-fitting, solution: $\Omega_\text{c0}$ can be larger provided that there is also a significant amount of DE, $\doc>0$. This DE must then have a large sound speed which ensures that its perturbation are erased already at large scales and the gravitational collapse of CDM proceeds more slowly at most subhorizon scales, according to the growth function~\eqref{eq:growth}.  Essentially all the points in the posterior presented in the right panel which have a large sound speed come from this alternative solution.}
\end{figure*}

In principle, one could have also varied the initial fluctuation amplitude $A_\text{S}$ which would have given a much weaker constraint for $\Omega_\text{c0}$, compatible with the CMB fits. However, the CMB measures the initial amplitude very well and therefore WL does not significantly correct the posterior for it. Thus the posteriors presented in Fig.~\ref{fig:wl} represent better the effective contribution of WL to the posterior of the combined data. We of course free all the parameters for the final combined fit.

\subsection{Combined constraints\label{sec:combined}}
For the combined analysis, we include all the CMB data in the Planck anisotropy power spectrum and the CFHTLeNS weak lensing shear. We free all the standard $\Lambda$CDM parameters as well as $\doc$ and $c_\text{s}^2$.

\begin{figure*}[t!]
\includegraphics[width=0.95\textwidth]{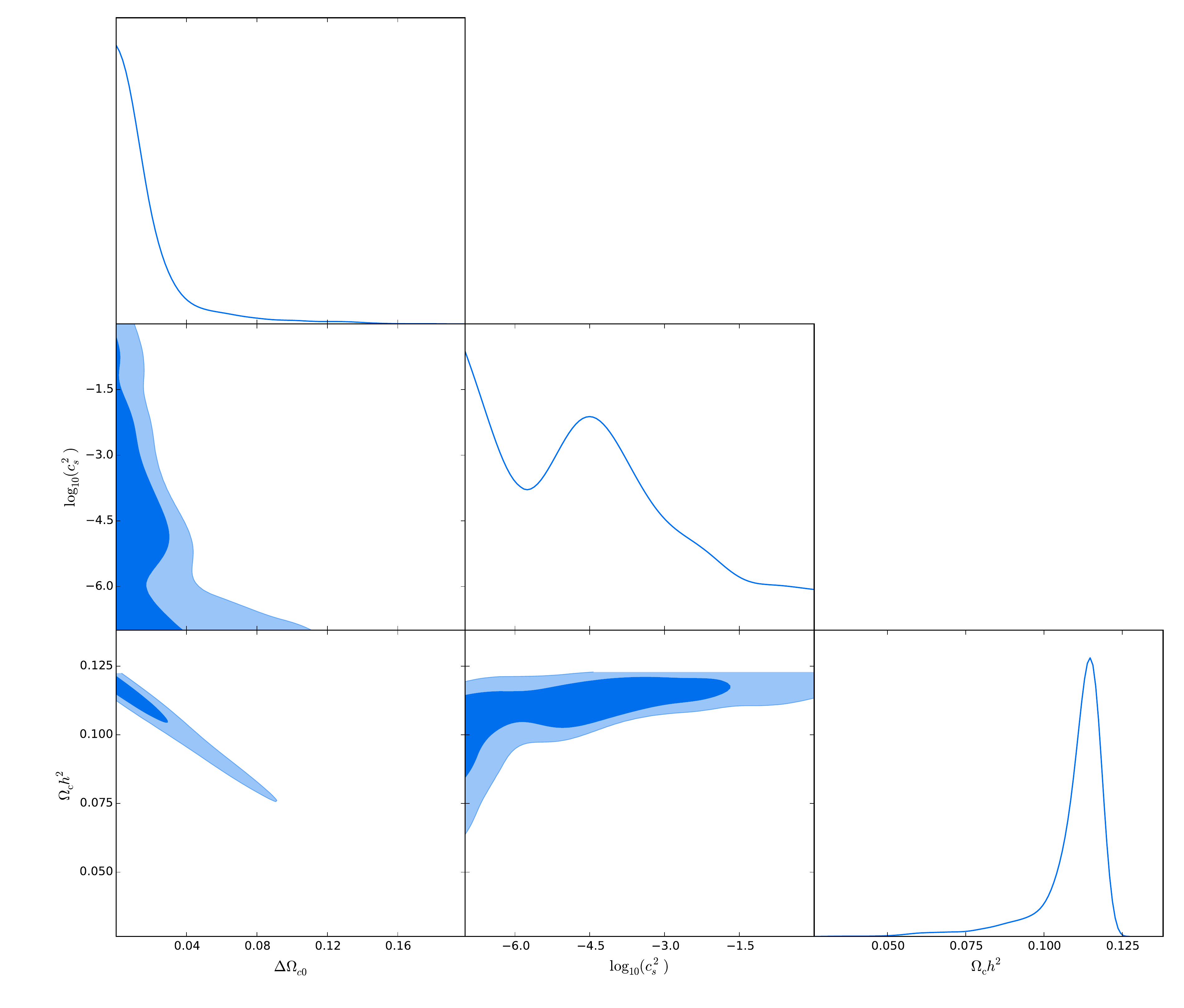}
\caption{\label{fig:combined} The constraints on $\Delta\Omega_\text{c0}$ and $c_s^2$ when using the combination of Planck CMB power-spectrum data and distance data (BAO+SNe), together with weak lensing shear data from CFHTLeNS. We have imposed a hard prior $c_\text{s}^2>10^{-7}$ in order to remove the infinite parameter space where our model is completely degenerate with $\Lambda$CDM for all the data under consideration. The data prefer a small admixture of DE with $\doc= 0.017 \pm 0.019$ and a sound speed of $\log_{10}(c_\text{s}^2)=-4.394 \pm 1.779$. This gives a cosmology which for the purposes of the CMB looks like the standard $\Lambda$CDM model, but the clustering is reduced at the scales probed by galaxy shear.}
\end{figure*}

Like in previous cases we find that a dark degeneracy is recovered in the combined fits for $c_\text{s}^2\lesssim10^{-7}$, i.e.~in the parameter range as determined by the WL measurement. This parameter range is equivalent to the concordance cosmology and offers no improvement in the combined fit.

However, there exists an alternative, superior solution already mentioned in Sec.~\ref{sec:WL}, which fits the combined data better than $\Lambda$CDM with a $\Delta\chi^2=2.4$ while containing two extra parameters. In order to discuss the posterior of this solution, we impose a prior to eliminate the part of infinite logarithmic parameter space where our model is completely degenerate with $\Lambda$CDM, cutting the chains at $c_\text{s}^2 >10^{-7}$. We show this part of the parameter space in Fig.~\ref{fig:combined}.

Given this cut, the combined data prefer $\Omega_\text{c0} = 0.239 \pm 0.020$ together with an admixture of $\doc= 0.017 \pm 0.019$ of dark energy with a sound speed of $\log_{10}(c_\text{s}^2)=-4.394 \pm 1.779$. The combined density fraction $\Omega_\text{c0}+\doc$ is consistent with the Planck 2013 best fit of $\Omega_\text{c0} = 0.265 \pm 0.011$. At small scales, the DE stops to cluster and recovers the preferred lensing density fraction.

We note that in this alternative solution, the amplitude of the CDM power spectrum is $\sigma_8=0.79 \pm 0.02$, which lies within 1$\sigma$ of the value as implied by clusters in Ref.~\cite{Ade:2013lmv}. However, the power spectrum is quite modified at small scales and the predictions for nonlinear structures might not be related to the value of $\sigma_8$ in the same manner as in $\Lambda$CDM.

\section{Discussion and Conclusions}

There is a weak but chronic tension between the CMB measurements of cosmological inhomogeneities and those obtained from the late universe, which consistently suggest that there is less power at late times than expected in $\Lambda$CDM. There are many systematic effects that could be playing a role in biasing our interpretations of the low-redshift universe (baryonic physics, intrinsic alignment, badly modeled nonlinear physics), but one should ask whether such a phenomenology, if it were to persist, could be evidence of some sort of dynamics in the dark energy/gravity sector.

The usual approach to solve this is to change the geometry of the universe at late times (i.e.~change $w$). However, it is difficult to create a large effect on perturbations without significantly altering distances. The constraints from BAOs and SNe are now tight enough that it is very difficult to create a large effect consistent with these background data and the CMB. Moreover, the CMB lensing and the WL are driven by the same physics and the geometrical kernels lie close together at lower redshifts, so that it is difficult to produce a significant discrepancy in the clustering seen by these two probes with the help of a purely time-dependent effect.

Taking account of the fact that the data in tension (CMB and WL/clusters) probe different scales, we advocate an alternative approach.  We keep the expansion history fixed and use the DE to make the CDM cluster differently at small and large scales. We have used a simple perfect fluid to model this effect. Yet this type of model is relatively poorly explored as a result of the focus on particular classes of parametrization of $w$ (CPL). We have instead  exploited the dark degeneracy to freely vary the CDM density fraction $\Omega_\text{c0}$ while keeping $H(z)$ fixed.

The sort of parametrization we have employed gives a non-negligible contribution of dark energy at early times. This does not deform the CMB in an unacceptable manner, because our model strongly violates the quasistatic approximation typically used in the DE/MG calculations. Perfect fluids at scales beyond their sound horizons evolve as dust, irrespective of their equation of state. Thus the growth rate at largest scales is completely determined by $H(z)$ and perturbation evolution in our model is  degenerate with $\Lambda$CDM at large enough scales. This is a known and generic result for any model with no anisotropic stress \cite{Bertschinger:2006aw} and exploited in the parametrization of Ref.~\cite{Hu:2007pj}.

That being said, CMB lensing, detected in the Planck power spectrum at 10$\sigma$, puts a very strong upper limit on the DE sound speed. Since the CMB lensing is consistent with $\Lambda$CDM, the DE must cluster as dust on the scales probed by it. Then there needs to be a rapid transition in the clustering properties, so that the CDM growth rate is reduced at the smaller scales probed by WL.  

We stress that despite the fact that the CMB is a high-redshift observable, it constrains our model mostly at low redshifts. In total, the combination the combination of CMB and WL data fit the model better than $\Lambda$CDM with a $\Delta\chi^2=2.4$. This solution has a $\Omega_\text{c0} = 0.239 \pm 0.020$ with the presence of a small admixture of DE $\doc= 0.017 \pm 0.019$. This keeps the CMB anisotropies unchanged while reducing the amplitude of the gravitational potential inside the Jeans horizon and bringing the two data sets together.

Our improved solution has reduced $\sigma_8=0.79 \pm 0.02$, which is closer to the result reported by Planck in 2013 from clusters ($\sigma_8=0.75\pm0.03$) \cite{Ade:2013lmv}. This decrement in power continues at small scales, which could in principle help alleviate some of the tension between N-body simulations of CDM and the observations (e.g.~the missing-satellites problem \cite{Klypin:1999uc} and Too Big to Fail \cite{BoylanKolchin:2011dk}). It is interesting to ask whether the nonlinear behavior of CDM in this scale-dependent model could be mapped from standard $\Lambda$CDM N-body simulation using a method such as that proposed in \cite{Mead:2014gia}. Since in our model at small scales gravity is not modified and there are no new screening effects, this might give a simple method for adjusting calibration in e.g.~Halofit for this subclass of DE models and therefore to predict small-scale lensing and cluster formation with a better accuracy than we can currently.

We have not used constraints from redshift-space distortions on the CDM growth rate $f\sigma_8$. On one hand, we find that the uncertainty in the most precise measurements to date \cite{Samushia:2013yga,Beutler:2013yhm, Chuang:2013hya} is still high enough not to affect the posteriors significantly in the vicinity of the best-fit region of our scenario. More importantly, we are dealing with a model in which an evolving scale-dependence of the matter power spectrum is a key feature. Interpreting the growth-rate data in such a scenario is somewhat subtle since the growth rates are scale-dependent \cite{Ballesteros:2008qk,Nesseris:2015fqa} and we will give this question the consideration it deserves in a separate work.

We should stress that we have exploited a completely generic feature of modified gravity. The response of the gravitational potential to the CDM perturbations depends on scale in all models apart from $\Lambda$CDM, whether because such a transition is already explicit in the quasistatic approximation (e.g.~$f(R)$ gravity where the Compton mass provides a scale \cite{Song:2006ej}, or see Ref.~\cite{Motta:2013cwa}) or because the quasistatic approximation fails beyond the Jeans scale (in the scenario discussed here; also see Ref.~\cite{Sawicki:2015zya}). This implies that a scale-dependent modification of the matter power spectrum is completely generic in dynamical dark energy/modified gravity models and typically will take place in a range of modes corresponding to the Jeans/Compton-mass scale today and mode corresponding to the Jeans scale at the time that the DE/MG started contributing to the gravitational potential significantly.

Observations of large-scale structure are performed not only around a particular redshift, but inside a particular range of scales probed by the survey. This scale-dependence can make the interpretation of measurements complicated, if the measurements are reported assuming a $\Lambda$CDM-like scale-independent behavior, instead of a rawer form closer to the observation.

The phenomenology of such ``cold dark energy'' \cite{Sapone:2010uy} models featuring low sound speeds as proposed here and exploiting a scale-dependent phenomenology, but in a more general context of effective field theory of dark energy \cite{Gubitosi:2012hu, Bloomfield:2012ff, Gleyzes:2013ooa, Gleyzes:2014qga, Bellini:2014fua} remains largely unexplored, but is necessary if we are to build a full understanding of the constraints data placed on the properties of the dark sector.

\begin{acknowledgments}
We are grateful to M.A.~Amin, K.C.~Chan, C.H.~Chuang, P.S.~Corasaniti, J.~Gleyzes, B.~Hu, A.~Mead, F.~Piazza, V.~Valkenburg and F.~Vernizzi. M.K.\ and S.N.\ acknowledge funding by the Swiss National Science Foundation. I.S.~is supported by the Maria Skłodowska-Curie Intra-European Fellowship Project ``DRKFRCS''. The numerical computations for our analysis were performed on the Baobab cluster at the University of Geneva.

The development of Planck was supported by: ESA; CNES and CNRS/INSU- IN2P3-INP (France); ASI, CNR, and INAF (Italy); NASA and DoE (USA); STFC and UKSA (UK); CSIC, MICINN and JA (Spain); Tekes, AoF and CSC (Finland); DLR and MPG (Germany); CSA (Canada); DTU Space (Denmark); SER/SSO (Switzerland); RCN (Norway); SFI (Ireland); FCT/MCTES (Portugal). A description of the Planck Collaboration and a list of its members, including the technical or scientific activities in which they have been involved, can be found at http://www.cosmos.esa.int/web/planck/planck-collaboration.
\end{acknowledgments}

\bibliographystyle{utcaps}
\bibliography{scales}

\end{document}